\documentstyle[aps,epsf,psfig]{revtex}  
\global\firstfigfalse 
\global\firsttabfalse  
\newcommand{\rms}{\rm\scriptstyle}
\newcommand{\rmt}{\rm\textstyle}
\newcommand{\nub}{\overline{\nu}}
\begin{document}        
\baselineskip 14pt

\title{A Measurement of $\mathbf{\sin^2\theta_W}$ in $\mathbf{\nu N}$ 
Scattering from NuTeV}

\author{
 G.~P.~Zeller$^{5}$,
 T.~Adams$^{4}$, A.~Alton$^{4}$, S.~Avvakumov$^{7}$,
 L.~de~Barbaro$^{5}$, P.~de~Barbaro$^{7}$, 
 R.~H.~Bernstein$^{3}$,
 A.~Bodek$^{7}$, T.~Bolton$^{4}$, J.~Brau$^{6}$, D.~Buchholz$^{5}$,
 H.~Budd$^{7}$, L.~Bugel$^{3}$, J.~Conrad$^{2}$, R.~B.~Drucker$^{6}$,
 J.~Formaggio$^{2}$,R.~Frey$^{6}$, J.~Goldman$^{4}$, M.~Goncharov$^{4}$,
 D.~A.~Harris$^{7}$, 
 R.~A.~Johnson$^{1}$, S.~Koutsoliotas$^{2}$,
 J.~H.~Kim$^{2}$, 
 M.~J.~Lamm$^{3}$, W.~Marsh$^{3}$, 
 D.~Mason$^{6}$, C.~McNulty$^{2}$, K.~S.~McFarland$^{7,3}$, 
 D.~Naples$^{4}$, P.~Nienaber$^{3}$, A.~Romosan$^{2}$, W.~K.~Sakumoto$^{7}$,
 H.~Schellman$^{5}$, M.~H.~Shaevitz$^{2}$, P.~Spentzouris$^{2}$, 
 E.~G.~Stern$^{2}$, B.~Tamminga$^{2}$, M.~Vakili$^{1}$, 
 A.~Vaitaitis$^{2}$, 
 V.~Wu$^{1}$, U.~K.~Yang$^{7}$ and J.~Yu$^{3}$}
\address{
 $^{1}$University of Cincinnati, Cincinnati, OH 45221 \\            
 $^{2}$Columbia University, New York, NY 10027 \\                   
 $^{3}$Fermi National Accelerator Laboratory, Batavia, IL 60510 \\  
 $^{4}$Kansas State University, Manhattan, KS 66506 \\              
 $^{5}$Northwestern University, Evanston, IL 60208 \\               
 $^{6}$University of Oregon, Eugene, OR 97403 \\                    
 $^{7}$University of Rochester, Rochester, NY 14627 } 

\maketitle              

\begin{abstract}        
The NuTeV experiment at Fermilab presents a determination of the electroweak 
mixing angle. High purity, large statistics samples of $\nu_\mu N$ and 
$\nub_\mu N$ events allow the use of the Paschos-Wolfenstein relation. 
This considerably reduces systematic errors associated with charm production 
and other sources. With Standard Model assumptions, this measurement of 
$\sin^{2}\theta_{W}$ indirectly determines the W boson mass to a precision 
comparable to direct measurements from high energy $e^+e^-$ and $p\bar{p}$ 
colliders. NuTeV measures
${\rmt sin^{2}\theta_{W}}^{({\rms on-shell)}}=0.2253\pm0.0019({\rmt stat.})\pm0.0010({\rmt syst.})$,
which implies $M_W=80.26\pm0.11~$ GeV.
\end{abstract} 

\section{Introduction}        

Neutrino scattering experiments have contributed to our understanding of 
electroweak physics for more than three decades. Early determinations of 
$\sin^{2}\theta_{W}$ served as the critical ingredient to the Standard 
Model's successful prediction of the W and Z boson masses. More precise 
investigations in the late 1980's set the first useful limits on the top 
quark mass. Just as early measurements contributed to the accurate 
predictions of the W, Z, and top quark masses before their direct 
observation, recent results from neutrino-scattering experiments 
combined with $e^+e^-$ and $p\bar{p}$ collider data likewise constrain 
the Higgs boson mass. \\

The measurement presented here represents the most precise determination 
of the electroweak mixing angle from neutrino-nucleon scattering to date. 
The result is a factor of two more precise than the previous most accurate 
$\nu$N measurement \cite{ccfr}.

\section{Methodology}

In deep inelastic neutrino-nucleon scattering, the weak mixing angle can
be extracted from the ratio of neutral current (NC) to charged current (CC)
total cross sections. Previous measurements relied on the Llewellyn-Smith
formula, which relates these ratios to $\sin^{2}\theta_{W}$ for neutrino 
scattering on isoscalar targets \cite{llewellyn-smith}:

\begin{equation}
R^{\nu} \equiv \frac{\sigma(\nu_{\mu}N\rightarrow\nu_{\mu}X)}
                 {\sigma(\nu_{\mu}N\rightarrow\mu^-X)}  
= \frac{1}{2}-\sin^2\theta_W+\frac{5}{9}(1+r)\sin^4\theta_W,
\end{equation}

\begin{equation}
R^{\nub} \equiv \frac{\sigma(\nub_{\mu}N\rightarrow\nub_{\mu}X)}
                 {\sigma(\nub_{\mu}N\rightarrow\mu^+X)}  
= \frac{1}{2}-\sin^2\theta_W+\frac{5}{9}(1+ \frac{1}{r})\sin^4\theta_W,
\end{equation}

where

\begin{equation}
r \equiv \frac{\sigma({\overline \nu}_{\mu}N\rightarrow\mu^+X)}
                {\sigma(\nu_{\mu}N\rightarrow\mu^-X)} \sim \frac{1}{2},  
\end{equation}

\noindent
The above equations are exact only for tree level scattering off 
an idealized isoscalar target composed of first generation light quarks. 
Corrections must be made for the non-isoscalar target, the heavy quark content 
of the nucleon, radiative effects, higher-twist processes, the longitudinal
structure function ($R_L$), and charm production. This last effect is  
most important. Unfortunately, previous determinations of $\sin^2\theta_W$ 
measured in this way were subject to the same charm production uncertainties 
(resulting from imprecise knowledge of the charm quark mass) that dominated 
the CCFR error \cite{ccfr}. This ultimately limited the precision of neutrino 
measurements of electroweak parameters. 

An alternate method for determining $\sin^2\theta_W$ that is much less 
dependent on the details of charm production and other sources of model 
uncertainty is derived from the Paschos-Wolfenstein quantity, $R^{-}$ 
\cite{pw}:

\begin{equation}
R^{-} \equiv \frac{\sigma(\nu_{\mu}N\rightarrow\nu_{\mu}X)-
                   \sigma(\nub_{\mu}N\rightarrow\nub_{\mu}X)}
                  {\sigma(\nu_{\mu}N\rightarrow\mu^-X)-  
                   \sigma(\nub_{\mu}N\rightarrow\mu^+X)}  
= \frac{R^{\nu}-rR^{\nub}}{1-r}=\frac{1}{2}-\sin^2\theta_W
\end{equation}

\noindent
Because $R^-$ is formed from the difference of neutrino and antineutrino
cross sections, almost all sensitivity to the effects of sea quark scattering 
cancels. This reduces the error associated with heavy quark production 
by roughly a factor of eight relative to the previous analysis. The 
substantially reduced uncertainties, however, come at a price. 
Unlike $R^{\nu}$, the ratio $R^-$ is more difficult to measure experimentally 
because neutral-current neutrino and antineutrino events have identical 
observed final states. The two samples can only be separated via
\emph{a priori} knowledge of the incoming neutrino beam type. 

\section{The Neutrino Beam and the NuTeV Detector}        

High-purity neutrino and antineutrino beams were provided by the Sign Selected 
Quadrupole Train (SSQT) at the Fermilab Tevatron during the 1996-1997
fixed target run. Neutrinos are produced from the decay of pions and kaons 
resulting from interactions of 800 GeV protons in a BeO target. Dipole magnets
immediately downstream of the proton target bend pions and kaons of specified 
charge in the direction of the NuTeV detector, while wrong-sign and neutral 
mesons are stopped in beam dumps. The resulting beam is almost purely 
neutrino or antineutrino depending on the selected sign of the parent mesons. 
The measured $\nub_{\mu}$ contamination in the $\nu_{\mu}$ beam is less than 
1/1000 and the $\nu_{\mu}$ contamination in the $\nub_{\mu}$ beam is less than
1/500. In addition, the beam is almost purely muon-neutrino with a small
contamination of electron neutrinos ($1.3\%$ in neutrino mode and 
$1.1\%$ in antineutrino mode).

\begin{figure}[ht]	
\centerline{\epsfxsize 3.5 truein \epsfbox{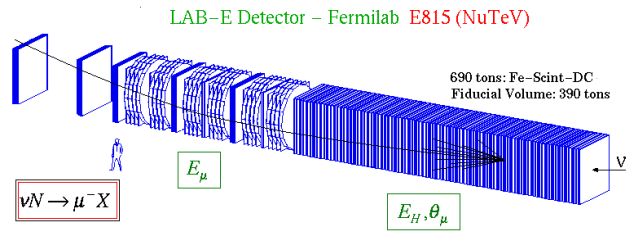}}   
\caption[]{The NuTev detector.}
\label{fig:labe}
\end{figure}

Neutrino interactions are then observed in the NuTeV detector, which is 
located approximately 1.5 km downstream of the proton target. 
The detector consists of an 18m long, 690 ton steel-scintillator target 
followed by an instrumented iron-toroid spectrometer (Figure \ref{fig:labe}). 
The target calorimeter is composed of 168 3 m x 3 m x 5.1 cm steel plates 
interspersed with liquid scintillation counters (spaced every two plates) 
and drift chambers (spaced every four plates). The scintillation counters 
provide triggering information as well as a determination of the longitudinal 
event vertex, event length and visible energy deposition. The mean position 
of hits in the drift chambers help establish the transverse event vertex. 
The toroidal spectrometer, which determines muon sign and momentum, is not 
directly used for this analysis.

\noindent
The detector was continuously calibrated through exposure to a wide energy 
range of test beam hadrons, electrons and muons delivered in a separate 
beamline during each accelerator cycle.

\section{Event Selection}

For inclusion in this analysis, events must deposit at least 20 GeV of visible
energy in the detector. This cut ensures full efficiency of the event 
trigger, proper vertex determination, and reduction of cosmic ray background 
events. Events must also have a vertex that lies within a 1.0 m box around 
the center of the detector which is at least 0.4 m of steel from the upstream 
end and 2.4 m of steel from the downstream end of the detector. This 
``box cut'' ensures shower containment and minimizes the contribution 
from $\nu_{e}$ interactions. The chosen longitudinal fiducial volume ensures 
both that the event is neutrino-induced and that a meaningful event length can
be measured. After all cuts, the total analysis sample consists of 1.3 million
neutrino and 0.3 million antineutrino events with a mean energy of 
approximately 125 GeV.

\section{Event Separation}        

\begin{figure}[hbt]
\begin{center}

\mbox{\psfig{figure=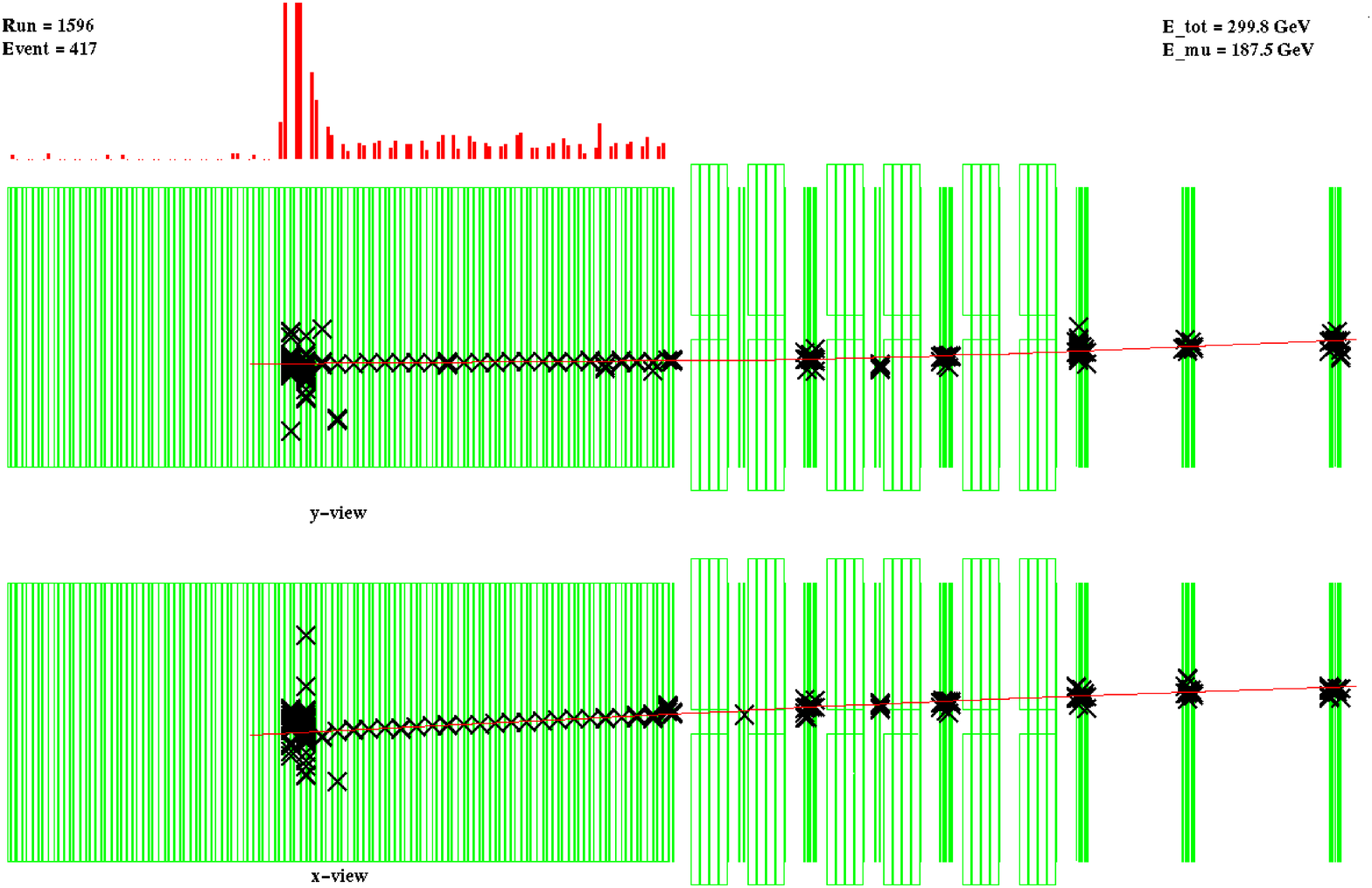,width=2.5in,bbllx=185bp,bblly=375bp,bburx=1012bp,bbury=770bp,clip=0,silent=0}
\epsfxsize=1.0in\epsfbox[0 -30 148 148]{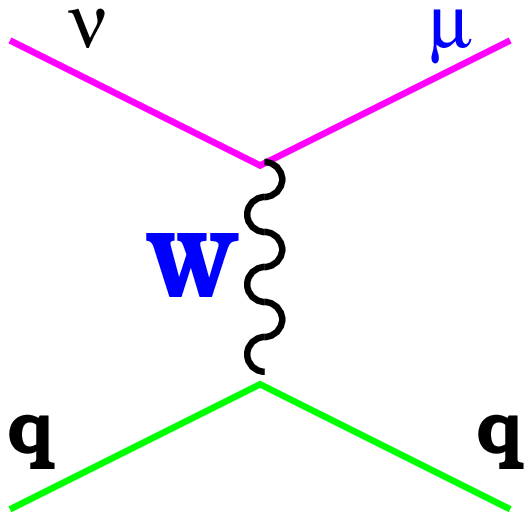}} 

\mbox{\psfig{figure=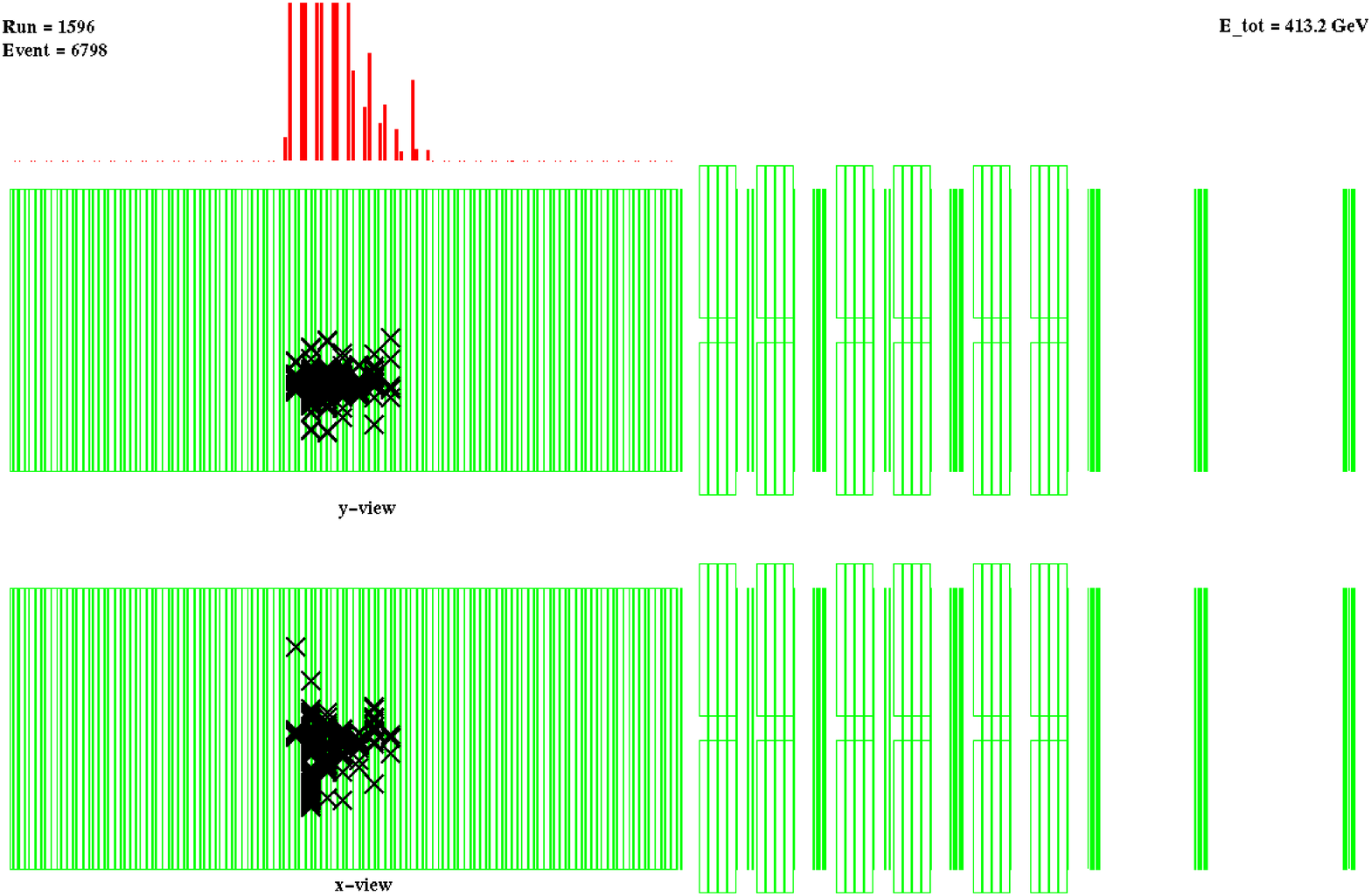,width=2.5in,bbllx=185bp,bblly=375bp,bburx=1012bp,bbury=770bp,clip=0,silent=0}
\epsfxsize=1.0in\epsfbox[0 -30 148 148]{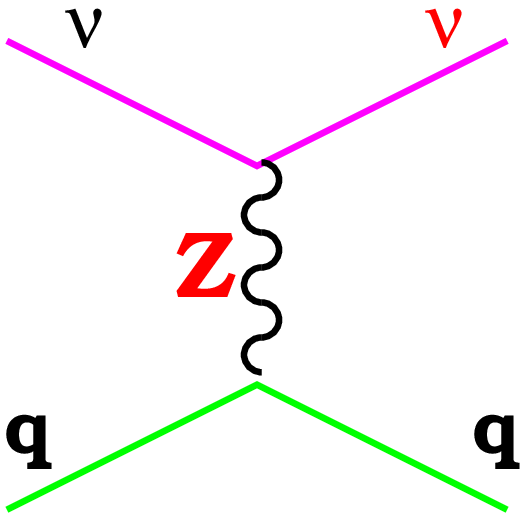}} 
\caption[]{Event displays of a NuTeV charged current candidate event (top) 
           and neutral current candidate (bottom).}
\label{fig:events}
\end{center}
\end{figure}

In order to measure $\sin^2\theta_W$, observed neutrino events must be 
separated into charged current (CC) and neutral current (NC) categories. 
Figure \ref{fig:events} shows typical CC and NC candidate events in the 
NuTeV detector.

Both CC and NC neutrino interactions initiate a cascade of hadrons in the 
target that is registered in both the scintillation counters and drift 
chambers. Muon-neutrino CC events are distinguished by the presence of a final 
state muon. The muon typically penetrates well beyond the hadronic shower and
deposits energy characteristic of a minimum-ionizing particle in a large 
number of scintillation counters. The muon track is clearly visible in the CC 
event at the top of Figure \ref{fig:events}. On the other hand, there is no 
final state muon in muon-neutrino NC events; the final state neutrino is 
invisible, so $\nu_{\mu}$ NC events can only be recognized by their hadronic 
shower.

Given the differing event topologies, one can statistically separate CC and NC 
interactions on the basis of event length (\emph{i.e.}, on the presence or 
absence of a muon in an event). The length of an event is based on the 
longitudinal energy deposition in the calorimeter and is simply the number of 
scintillation counters spanned by the event. Events with a long length 
(spanning $>$ 20 counters) are identified as CC candidates and events with a 
short length (spanning $\leq$ 20 counters) are identified as NC candidates. 
The experimental quantity that is measured in both modes is the ratio:

\begin{equation}
{\rmt R_{meas}} = \frac{\# \hspace{0.05in} \rmt SHORT \hspace{0.05in} events}
                       {\# \hspace{0.05in} \rmt LONG \hspace{0.05in} events}
              = \frac{\# \hspace{0.05in} \rmt NC  \hspace{0.05in} candidates}
                     {\# \hspace{0.05in} \rmt CC \hspace{0.05in} candidates}
\end{equation}

Neither sample is of course entirely pure. Long events are predominantly 
$\nu_{\mu}$ CC interactions but contain small contaminations from both 
$\nu_{\mu}$ NC events where the hadronic shower fluctuates to long lengths, 
as well as muons which are produced in upstream neutrino interactions in the
shielding and later undergo a hard bremsstrahlung interaction in the detector. 
Short events are primarily $\nu_{\mu}$ NC interactions but also contain 
contributions from low energy $\nu_{\mu}$ CC events in which the muon ranges 
out, wide angle $\nu_{\mu}$ CC events in which the muon exits the side of 
the detector, $\nu_{e}$ CC events and cosmic rays. The ratios 
({$\rmt R_{meas}$}) 
of short to long events measured from the NuTeV data, are 0.4198 $\pm$ 0.0008 
in the neutrino beam and 0.4215 $\pm$ 0.0017 in the antineutrino beam. 
A Standard Model value of $\sin^2\theta_W$ can be directly extracted from 
these measured ratios by using a detailed Monte Carlo simulation of the 
experiment. The Monte Carlo must include the integrated neutrino fluxes, 
the neutrino cross section as well as a detailed description of the NuTeV 
detector. The following sections will discuss these three components.

\section{The muon-neutrino and electron-neutrino Fluxes}
The Monte Carlo simulation requires neutrino flux information as input. 
In particular, a precise determination of the electron neutrino contamination
in the beam is essential. The measured short to long ratios are directly 
impacted by the presence of electron neutrinos in the data sample because
$\nu_{e}$ charged current interactions, which usually lack an energetic 
muon in the final state, are almost always identified as neutral current 
interactions in the detector. 

\begin{figure}[ht]	
\centerline{\epsfxsize 6.5 truein \epsfbox{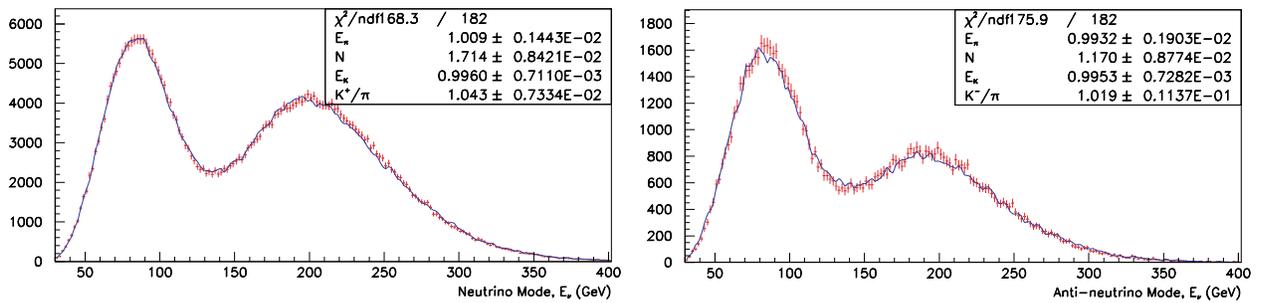}}   
\caption[]{$\nu_{\mu}$ and $\nub_\mu$ energy spectra comparing 
           data (points) and the tuned beam Monte Carlo (histogram).}
\label{fig:flux}
\end{figure}

The electron neutrino flux is estimated using a detailed beam Monte Carlo. 
Most of the observed $\nu_{e}$'s result from 
$K^{\pm}\rightarrow \pi^0e^{\pm} \nu_{e}$\hspace{-0.15in}$^{^{(-)}}$ decay. 
The beam simulation can be tuned to describe $\nu_{e}$ and 
$\nub_{e}$ production from charged kaon decay with high accuracy 
given that the $K^{\pm}$ decay contribution is well constrained from 
measurements of the observed $\nu_\mu$ and $\nub_\mu$ fluxes 
(as shown in Figure \ref{fig:flux}). Because of the precise alignment of the 
beamline elements and the low acceptance for neutral particle propagation 
in the SSQT, the largest uncertainty in the calculated electron neutrino flux 
results from the 1.5\% uncertainty in the 
$K^{\pm}\rightarrow \pi^0e^{\pm} \nu_{e}$\hspace{-0.15in}$^{^{(-)}}$ branching
ratio. The result is a factor of three reduction in the electron-neutrino flux 
uncertainty when compared to CCFR.

\hspace{0.1in}

\section{Cross Section Model}

Neutrino-nucleon deep-inelastic scattering processes are simulated using a 
leading order cross section model. The cross section model incorporates 
leading order parton momentum distributions measured using the same target 
and cross section model as NuTeV \cite{ccfr_sf}. Small modifications adjust
the parton densities to produce the inherent up-down quark asymmetry 
consistent with muon scattering \cite{muon} and Drell-Yan \cite{drell} data.
A leading order analysis of dimuon events in CCFR \cite{dimuon} provides
the shape and magnitude of the strange sea. Mass suppression from 
charged-current charm production is modeled using a leading order 
slow-rescaling formalism whose parameters are measured from the same 
high-statistics dimuon sample. Electroweak and QED radiative corrections 
to the scattering cross sections are applied using radiative correction 
routines supplied by D. Yu. Bardin \cite{bardin}. Higher twist contributions 
are quantified using a Vector-Dominance-Model as implemented by Pumplin 
\cite{pumplin} and constrained by lepto-production data \cite{lepto}. 
A global analysis performed by L. Whitlow \cite{whitlow} provides a
parameterization of the longitudinal structure function, $R_{L}$.

\section{Detector Response}

The Monte Carlo must also accurately simulate the response of the detector to 
the product of neutrino interactions in the target. The critical detector 
parameters that must be modeled are the calorimeter response to muons, 
measurement of the neutrino interaction vertex, and the range of hadronic 
showers in the calorimeter. Precise determination of the various detector 
effects are made possible through extensive use of both neutrino data and 
large samples of calibration beam data. The efficiency, noise, and effective 
size of the scintillation counters are measured using neutrino data or 
test beam muons. Vertex finding resolutions and biases are studied using 
neutrino data combined with information from a detailed GEANT-based simulation
of the detector. Test beam pion data provides information on hadronic shower 
lengths. 

\section {Event Length Distributions}

\begin{figure}[ht]	
\centerline{\epsfxsize 4.5 truein \epsfbox{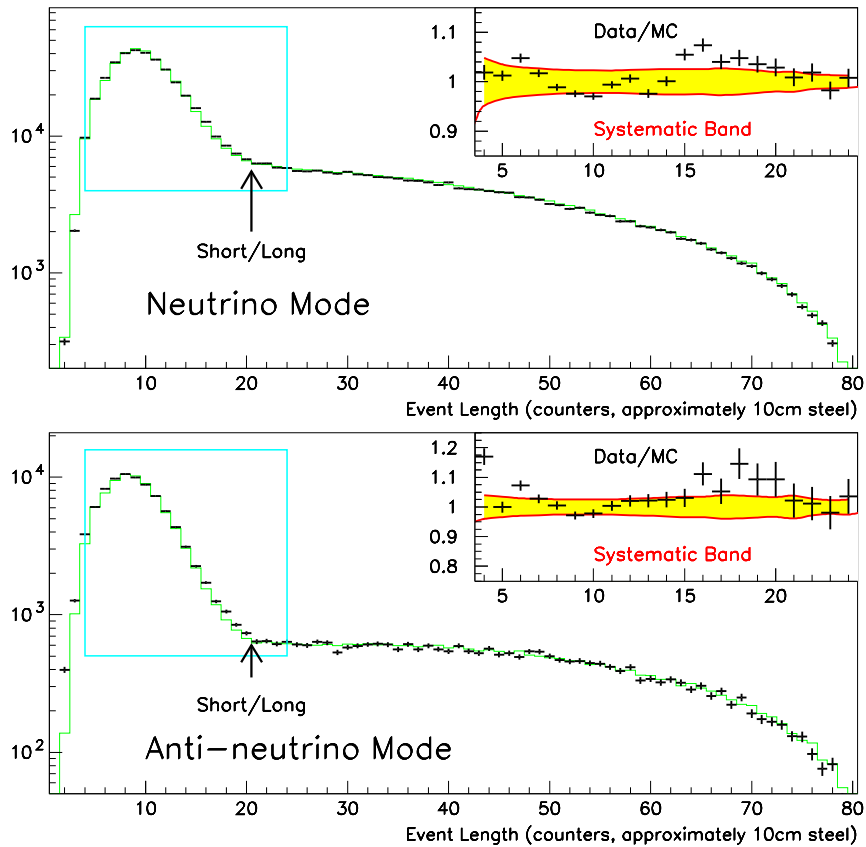}}
\vspace{0.3in}   
\caption[]{Event length comparisons for neutrino and antineutrino events.
           The solid curve is the Monte Carlo prediction. The neutral current 
           to charged current event separation (indicated by the arrow) is 
           made at a length of 20 counters, approximately 2m of steel.}
\label{fig:length}
\end{figure}

An important test of the Monte Carlo is its ability to predict the length
distribution of events in the detector. Figure \ref{fig:length} shows  
event length distributions of the final data sample compared to the Monte 
Carlo prediction assuming the best-fit $\sin^2\theta_W$ (see next section). 
Events reaching the toroid, which comprise about 80\% of the CC sample, have 
been left out for clarity but are included in the normalization of the Monte 
Carlo to the data. Data and MC agree well in each beam mode.

\hspace{0.1in}
\hspace{0.1in}

\section{Extraction of $\mathbf{\sin^2\theta_W}$}        

Given separate high-purity neutrino and antineutrino data sets, NuTeV 
measures the following linear combination of $R^{\nu}$ and $R^{\nub}$:

\begin{equation}
     R^{-} = R^{\nu}-x R^{\nub},
\end{equation}

\noindent
where x is chosen to be 0.5316. This value is obtained using Monte Carlo to 
minimize the dependence of $R^{-}$ on the charm quark mass. This approach 
explicitly minimizes uncertainties related to the suppression of charm quark 
production, largely eliminates uncertainties related to sea quark scattering, 
and reduces many of the theoretical and detector uncertainties common to both 
the neutrino and antineutrino samples. The single remaining free parameter in 
the Monte Carlo, $\sin^2\theta_W$, is then varied until the model calculation 
of $R^{-}$ agrees with what is measured in the data. The preliminary result 
from the NuTeV data sample for $M_{\rms top}$=175 GeV and 
$M_{\rms Higgs}$=150 GeV is:

\begin{equation}
    \sin^2\theta_W^{({\rms on-shell)}}=0.2253\pm0.0019({\rmt stat.})\pm0.0010({\rmt syst.})
\end{equation}

\begin{figure}[ht]	
\mbox{
\begin{minipage}{0.4\textwidth}
\centerline{\epsfxsize 2.5 truein \epsfbox{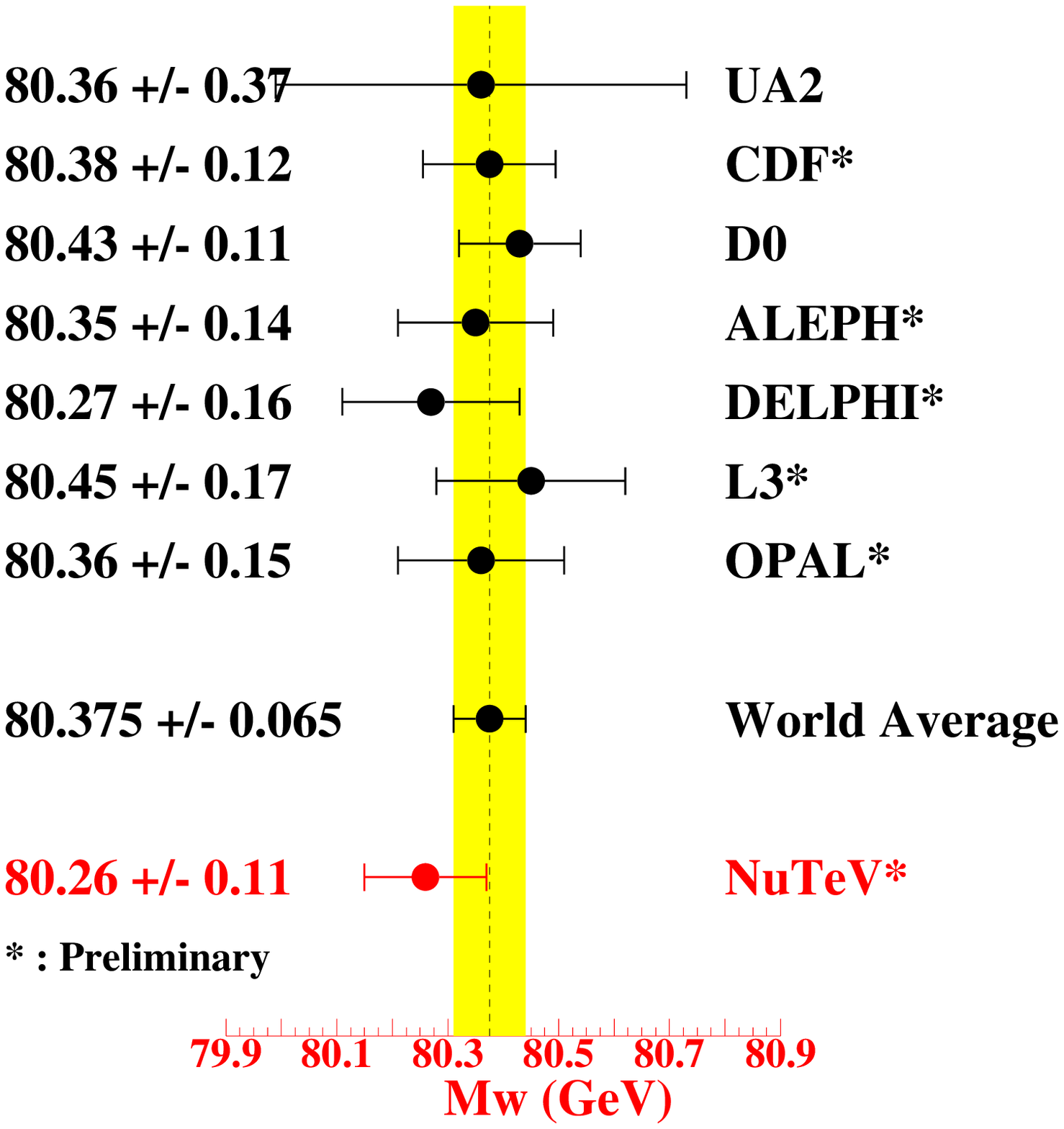}}   
\caption[]{Direct W boson mass measurements compared with this result
           (at the time of DPF99).}
\label{fig:mw}
\end{minipage}\hspace*{0.02\textwidth}	
\begin{minipage}{0.4\textwidth}
\centerline{\epsfxsize 2.5 truein \epsfbox{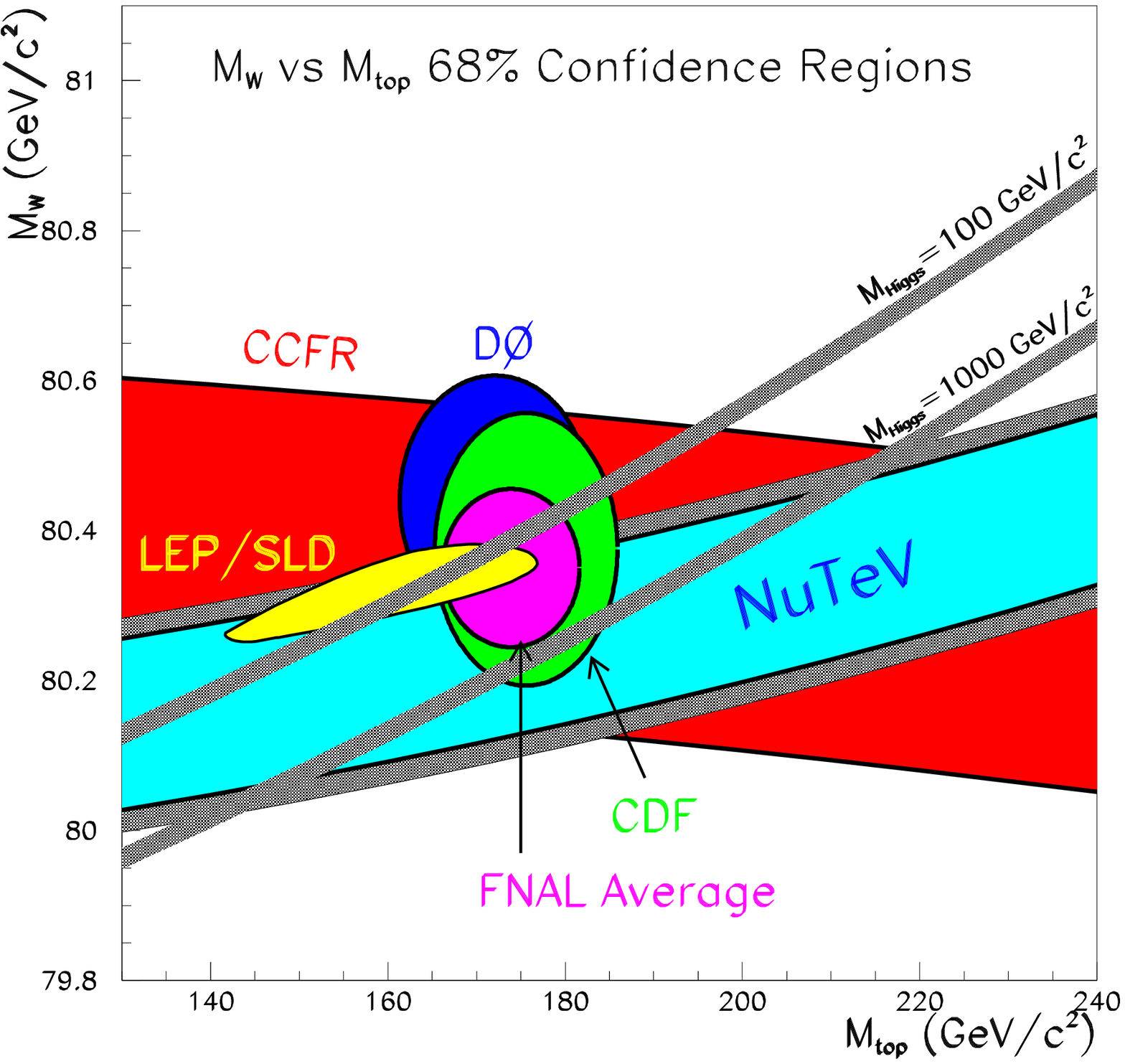}}   
\caption[]{Experimental constraints presented on the $M_{W}$-$M_{\rms top}$ 
           plane (at the time of DPF99). The two narrow lines indicate 
           the Standard Model predictions for $M_{\rms Higgs}$=100 and 
           1000 GeV.}
\label{fig:mt-mw}
\end{minipage}
}
\end{figure}

\noindent
Leading terms in the one-loop electroweak radiative corrections \cite{bardin} 
to the W and Z self-energies produce a small residual dependence of our 
result on $M_{\rms top}$ and $M_{\rms Higgs}$. We explicitly quote the 
dependence of our measurement on the top and Higgs masses. The effect 
is small given the existing uncertainty on the top quark mass and the 
logarithmic dependence on the Higgs mass. For example, the total change in 
$\sin^2\theta_W$ from varying the Higgs mass from 100 GeV to 1 TeV is only 
0.0005 (as is indicated by the shaded gray bands outlining the NuTeV result 
in Figure \ref{fig:mt-mw}).

\begin{equation}
  \delta \sin^2\theta_W=
         -0.00435\left[\left(\frac{M_{\rms top}}{175 GeV}\right)^{2}-1\right] 
         +0.00048\log \frac{M_{\rms Higgs}}{150 GeV}. \nonumber
\end{equation}

\noindent
Having chosen the convention, 
$\sin^2\theta_W^{({\rms on-shell)}}\equiv 1- \frac{M_{W}^2}{M_{Z}^2}$, 
where $M_{W}$ and $M_{Z}$ are the physical gauge boson masses, 
our result implies:

\begin{equation}
    M_{W}=80.26\pm0.10({\rmt stat.})\pm0.05(\rmt syst.)
         =80.26\pm0.11 GeV
\end{equation}

\noindent
A comparison of this result with direct measurements of $M_W$ is shown in 
Figure \ref{fig:mw}. Our measurement is in good agreement with Standard Model 
expectations and is consistent with the most recent measurements from W and Z 
production as well as from other neutrino experiments 
(Figure \ref{fig:mt-mw}). 

\noindent

\begin{figure}[ht]	
\mbox{
\begin{minipage}{0.4\textwidth}
\centerline{\epsfxsize 2.5 truein \epsfbox{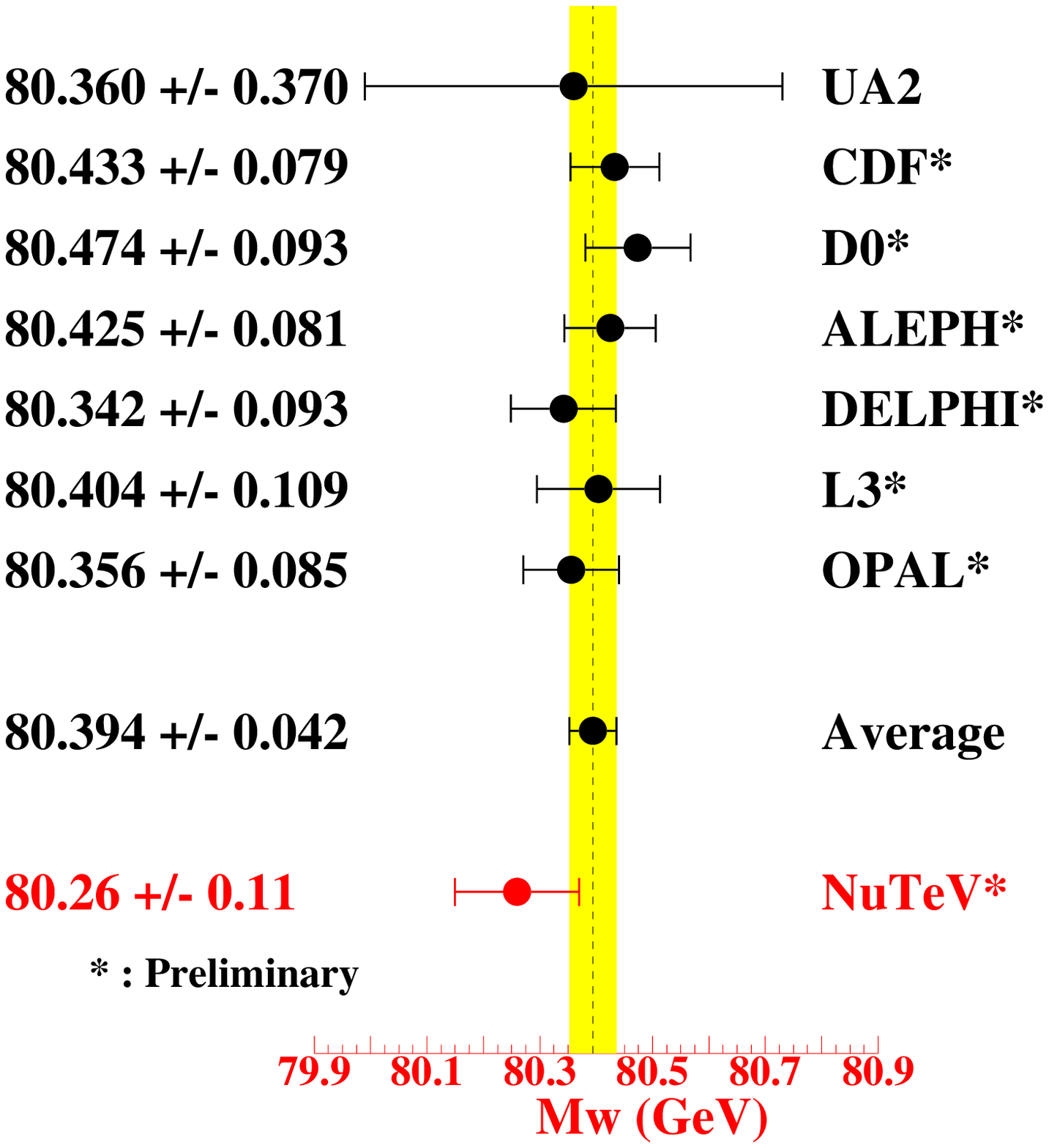}}   
\caption[]{Direct W boson mass measurements compared with this result (updated
           to include new measurements presented at EW Moriond99).}
\label{fig:mw99}
\end{minipage}\hspace*{0.02\textwidth}	
\begin{minipage}{0.4\textwidth}
\centerline{\epsfxsize 2.5 truein \epsfbox{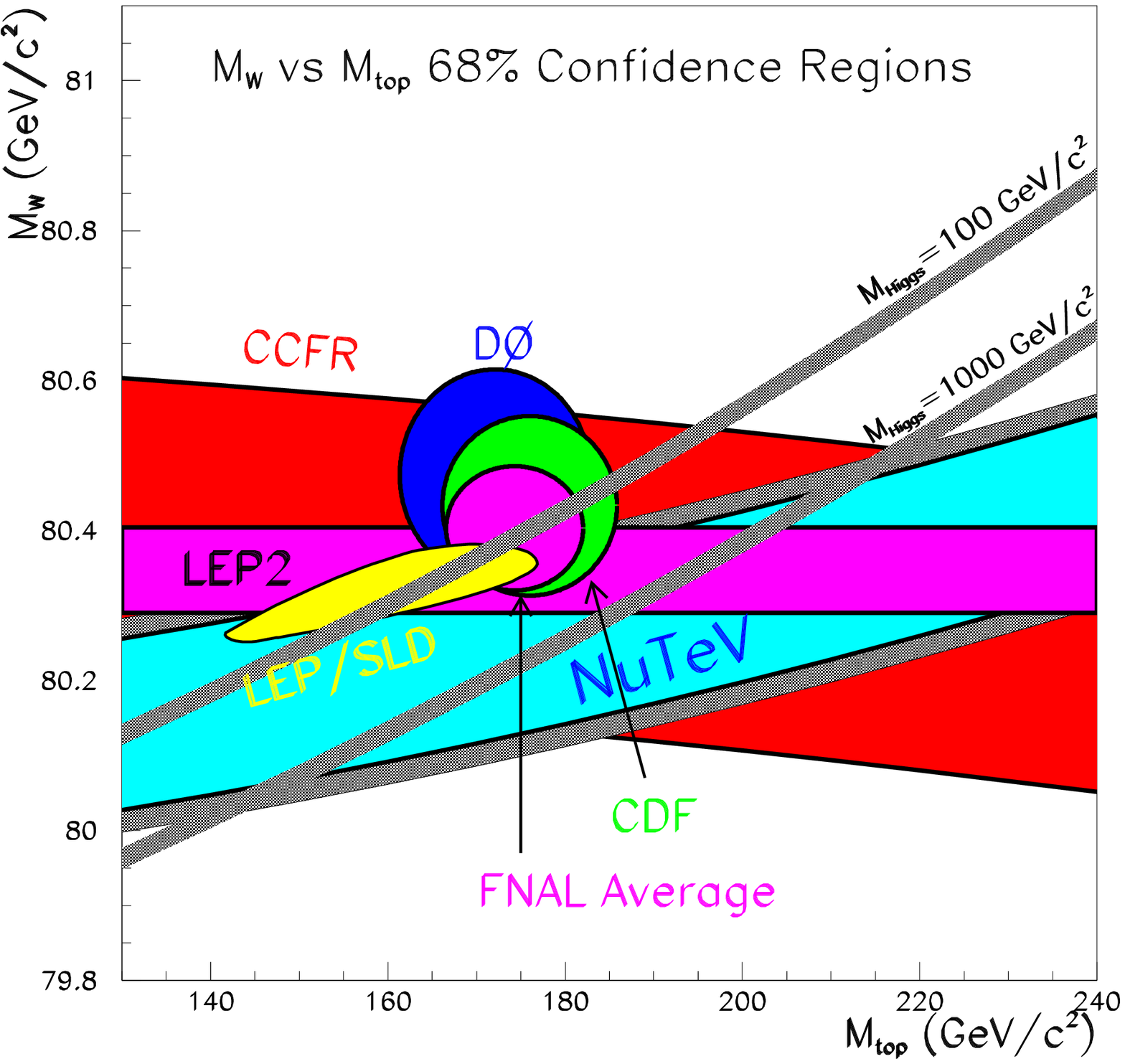}}   
\caption[]{Experimental constraints presented on the $M_{W}$-$M_{\rms top}$ 
          plane (updated to include new measurements presented at EW 
          Moriond99).}
\label{fig:mt-mw99}
\end{minipage}
}
\end{figure}

If these plots are now updated to include new measurements from 
CDF \cite{cdf}, D\O \cite{dzero} and LEP \cite{lep} that were presented at 
Electroweak Moriond in March of this year, it is interesting to note that 
tighter constraints on the Higgs mass are beginning to emerge 
(Figure \ref{fig:mt-mw99}).

\section{Conclusions}
NuTeV has successfully completed its data-taking and has extracted a 
preliminary value of $\sin^2\theta_W$. The precision of this result is 
a factor of two improvement over previous measurements in $\nu N$ 
scattering because of reduced systematics associated with measuring the 
Paschos Wolfenstein ratio, $R^{-}$. Interpreted within the framework of
the Standard Model, this result is equivalent to a determination of the 
W mass and is consistent with direct measurements of $M_{W}$.


\begin{references} 
\bibitem{ccfr} K.S. McFarland, {\em et al.}, Eur. Phys. Jour. {\bf C31}, 509 (1998).
\bibitem{llewellyn-smith} C.H. Llewellyn Smith, Nucl. Phys. {\bf B228}, 205 (1983).
\bibitem{pw} E.A. Paschos and L. Wolfenstein, Phys. Rev. {\bf D7}, 91 (1973).
\bibitem{ccfr_sf} W.G. Seligman, {\em et al.}, Phys. Rev. Lett. {\bf 79}, 1213 (1997).
\bibitem{muon} M. Arneodo, {\em et al.}, Nucl. Phys., {\bf B487}, 3 (1997)
\bibitem{drell} E.A. Hawker, {\em et al.}, Phys. Rev. Lett. {\bf 80}, 3715 (1998).
\bibitem{dimuon} S.A. Rabinowitz, {\em et al.}, Phys. Rev. Lett {\bf 70}, 134 (1993).
\bibitem{bardin} D. Yu. Bardin, V.A. Dokuchaeva, JINR-E2-86-260 (1986).
\bibitem{cdf} Y. Kim for the CDF collaboration, proceedings of La Thuile 1999.
\bibitem{dzero} A. Kotwal for the D\O  collaboration, D\O  Note 3544 (1999).
\bibitem{lep} LEP Electroweak Working Group, LEPEWWG/WW/99-01, revised (1999).
\bibitem{lepto} M. Virchaux and A. Milsztajn, Phys. Lett. {\bf B274}, 221 (1992).
\bibitem{pumplin} J. Pumplin, Phys. Rev. Lett. {\bf 64}, 2751 (1990). 
\bibitem{whitlow} L.W. Whitlow, SLAC-REPORT-357, 109 (1990).

\end{references}
\end{document}